\documentclass[conference]{IEEEtran}
\IEEEoverridecommandlockouts
\usepackage{cite}
\usepackage[linesnumbered,ruled,vlined]{algorithm2e}
\usepackage{amsmath,amssymb,amsfonts}
\usepackage{algorithmic}
\usepackage{graphicx}
\usepackage{textcomp}
\usepackage{xcolor}
\usepackage{url}
\def\BibTeX{{\rm B\kern-.05em{\sc i\kern-.025em b}\kern-.08em
    T\kern-.1667em\lower.7ex\hbox{E}\kern-.125emX}}

\begin{document}

% \title{SynDCIM: A Performance-Aware Digital Computing-in-Memory Compiler with Multi-Spec-Oriented Subcircuit Synthesis\vspace{-5mm}}

\title{\textcolor{black}{SynDCIM: A Performance-Aware Digital Computing-in-Memory Compiler with Multi-Spec-Oriented Subcircuit Synthesis\vspace{-5mm}}}

% \title{\huge SynDCIM: A Performance-Aware Digital Computing-in-Memory Compiler with Multi-Spec-Oriented Subcircuit Synthesis}

% \author{\IEEEauthorblockN{Jingyu He\textsuperscript{1}, Ziyang Shen\textsuperscript{2}, Fengshi Tian\textsuperscript{1}, Jinbo Chen\textsuperscript{2}, \\
% Jie Yang\textsuperscript{2}, Mohamad Sawan\textsuperscript{2}, Tim Cheng\textsuperscript{1}, Paul Bogdan\textsuperscript{3}, Chi-Ying Tsui\textsuperscript{1}}

% % \IEEEauthorblockA{\textsuperscript{1}Hong Kong University of Science and Technology\\
% % \textsuperscript{2} Westlake University \\
% % \textsuperscript{3} University of Southern California\\
% % }
% \vspace{-10mm}
% }

 \author{\IEEEauthorblockN{Kunming Shao\textsuperscript{1,2,*}, Fengshi Tian\textsuperscript{1,2,*,†}, Xiaomeng Wang\textsuperscript{1,2}, Jiakun Zheng\textsuperscript{1,2}, Jia Chen\textsuperscript{1,2},\\ Jingyu He\textsuperscript{1,2}, Hui Wu\textsuperscript{3}, Jinbo Chen\textsuperscript{3}, Xihao Guan\textsuperscript{1,2}, Yi Deng\textsuperscript{1,2}, Fengbin Tu\textsuperscript{1,2}, Jie Yang\textsuperscript{3},\\ Mohamad Sawan\textsuperscript{3}, Tim Kwang-Ting Cheng\textsuperscript{1,2}, Chi-Ying Tsui\textsuperscript{1,2}}

 \thanks{*Both authors contributed equally.}

 \thanks{†Corresponding author is Fengshi Tian (fengshi.tian@connect.ust.hk).}

 \thanks{This research was supported by ACCESS – AI Chip Center for Emerging Smart Systems, sponsored by InnoHK funding, Hong Kong SAR.}

 \IEEEauthorblockA{
 \textsuperscript{1}The Hong Kong University of Science and Technology, Hong Kong SAR, China\\
 \textsuperscript{2}AI Chip Center for Emerging Smart Systems (ACCESS), Hong Kong SAR, China\\
 \textsuperscript{3}Westlake University, Hangzhou, China
 }

%\IEEEauthorblockA{
% \textsuperscript{1}The Hong Kong University of Science and Technology,
% \textsuperscript{2}AI Chip Center for Emerging Smart Systems (ACCESS),
% \textsuperscript{3}Westlake University
% }
 
\vspace{-10mm}
}

\maketitle

\begin{abstract}
Digital Computing-in-Memory (DCIM) is an innovative technology that integrates multiply-accumulation (MAC) logic directly into memory arrays to enhance the performance of modern AI computing. However, the need for customized memory cells and logic components currently necessitates significant manual effort in DCIM design. Existing tools for facilitating DCIM macro designs struggle to optimize subcircuit synthesis to meet user-defined performance criteria, thereby limiting the potential system-level acceleration that DCIM can offer. To address these challenges and enable the agile design of DCIM macros with optimal architectures, we present SynDCIM — a performance-aware DCIM compiler that employs multi-spec-oriented subcircuit synthesis. SynDCIM features an automated performance-to-layout generation process that aligns with user-defined performance expectations. This is supported by a scalable subcircuit library and a multi-spec-oriented searching algorithm for effective subcircuit synthesis. The effectiveness of SynDCIM is demonstrated through extensive experiments and validated with a test chip fabricated in a 40nm CMOS process. Testing results reveal that designs generated by SynDCIM exhibit competitive performance when compared to state-of-the-art manually designed DCIM macros.
\end{abstract}


\section{Introduction}

As data movement between memory and processing units significantly contributes to energy consumption in AI computing, Computing-in-Memory (CIM) techniques have emerged as a promising solution for efficient AI operations. Among the various CIM architectures, SRAM-based Digital Computing-in-Memory (DCIM) designs demonstrate notable scalability and robustness against process, voltage, and temperature variations. These architectures integrate digital multiplication-and-accumulation (MAC) logic directly within SRAM memory arrays, facilitating distributed computing capabilities.

While numerous DCIM macro designs have shown advantages from technology scaling \cite{chih202116, fujiwara20225, mori20234nm, fujiwara202434}, their formulation often necessitates a tailored coupling of memory cells and MAC logic. Additionally, the inclusion of read-out peripherals to accommodate bit-configurable integer and floating-point precisions requires substantial manual design effort. The absence of comprehensive design automation tools for DCIM macros restricts their adaptability for diverse AI applications and complicates their integration into modern digital very-large-scale integration (VLSI) workflows. This highlights the urgent need for DCIM compilers that can streamline the generation of DCIM macros from user-defined specifications to final circuit layouts.

\begin{table}[t]
\caption{\textcolor{black}{Comparison with Emerging CIM Compilers.}}
\centering
\includegraphics[scale=0.6]{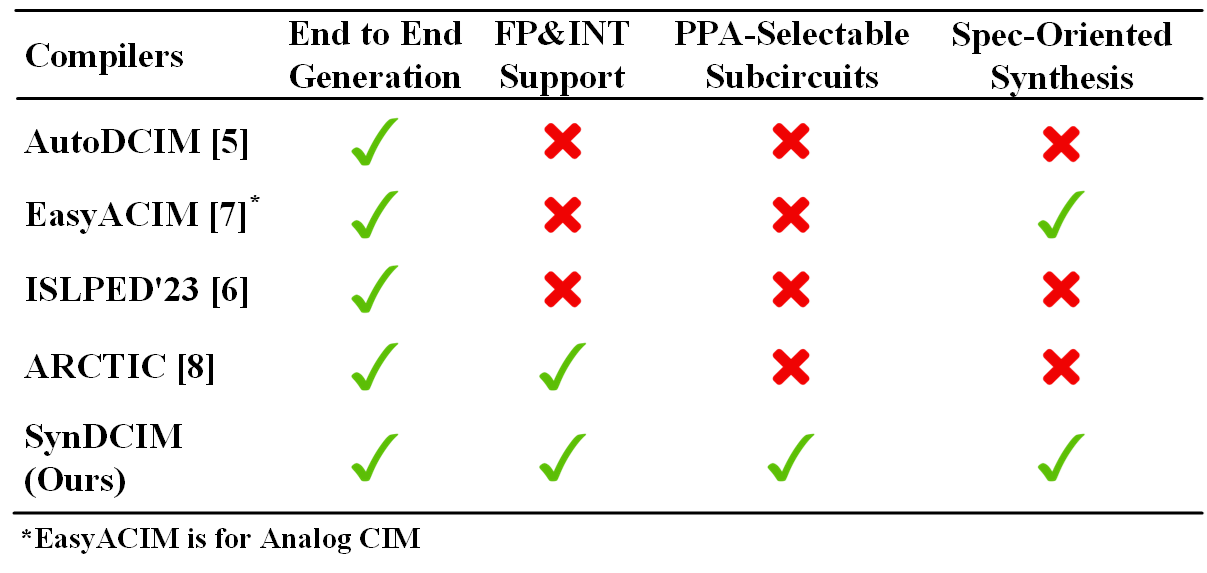}
\vspace{-3mm}
\label{tab:comparison_dcim_compiler} 
\end{table}

Inspiration from standardized SRAM macro generation has led to efforts in developing DCIM compilers capable of automatic macro generation. These compilers typically assemble template-based cell layouts to create an array \cite{chen2023autodcim, lanius2023automatic, zhang2024easyacim}. Some research has focused on parameterizing peripheral components to support various data formats \cite{zhang2024arctic}. Despite these advancements, a significant gap remains between the expected performance of DCIM systems and the current macro generation processes. A notable innovation in modern digital VLSI design tools is the integration of synthesis methods, such as the Synopsys Design Compiler, which converts behavior-level logic descriptions into gate-level netlists while optimizing for user-defined performance metrics like power, timing, and area. However, existing DCIM compilers often overlook the optimization of multiple performance specifications—such as energy efficiency, throughput, and area efficiency—during the synthesis of memory arrays and subcircuits, leading to suboptimal performance in the resulting DCIM macros.

Different AI applications—such as vision, language processing, and robotics—along with various acceleration scenarios, including wearable devices, mobile platforms, and cloud computing, necessitate distinct performance optimizations for DCIM macros. This requirement calls for the development of a scalable subcircuit library and corresponding synthesis algorithms. The subcircuit library should encompass fundamental logic components of a DCIM macro, including SRAM cells, MAC logic, and adders, while also modelling their performance characteristics. These models will assist synthesis algorithms in identifying optimal subcircuit combinations that meet the desired performance metrics of the DCIM macro. Such specialized electronic design automation (EDA) tools are termed performance-aware DCIM compilers; however, none of the existing solutions have successfully achieved this goal.

% \begin{table*}[htb]
% \caption{\textcolor{black}{Comparison with State-of-the-Art Works on MIT-BIH Dataset.}}
% \centering
% \includegraphics[scale=0.51]{figs/table3.png}
% \label{tab:comparison_all} 
% \end{table*}

To address the diverse execution requirements of various DCIM-based AI applications, this paper introduces SynDCIM, the first performance-aware DCIM compiler featuring multi-spec-oriented subcircuit synthesis (Table \ref{tab:comparison_dcim_compiler}). SynDCIM takes user-defined performance metrics as inputs, automatically determines the optimal DCIM architecture, and generates a complete layout for fabrication or further system integration. The main contributions of this work are summarized as follows:
\begin{itemize}
\item \textbf{SynDCIM Development}: We present SynDCIM, an automated performance-to-layout compiler that produces optimal DCIM macro designs aligned with user-defined performance expectations. SynDCIM accommodates flexible data precisions, scalable array parameters, and multiple-specification-oriented optimizations.
\item \textbf{Comprehensive Subcircuit Library}: SynDCIM includes a scalable subcircuit library featuring customizable components and performance look-up tables, providing essential building blocks for DCIM macro synthesis and layout generation.
\item \textbf{Multi-Spec-Oriented Searching Algorithm}: SynDCIM employs an innovative multi-spec-oriented searching algorithm that synthesizes subcircuits to identify optimal combinations that satisfy the specified performance criteria.
\item \textbf{Experimental Validation}: We validate the effectiveness of SynDCIM using a silicon-verified DCIM macro and demonstrate its superiority over existing solutions, while also evaluating performance trade-offs within constrained design spaces.
\end{itemize}

\section{Preliminaries of DCIM}

\subsection{DCIM Architecture}

\begin{figure}[tp]
\centerline{\includegraphics[width=\columnwidth]{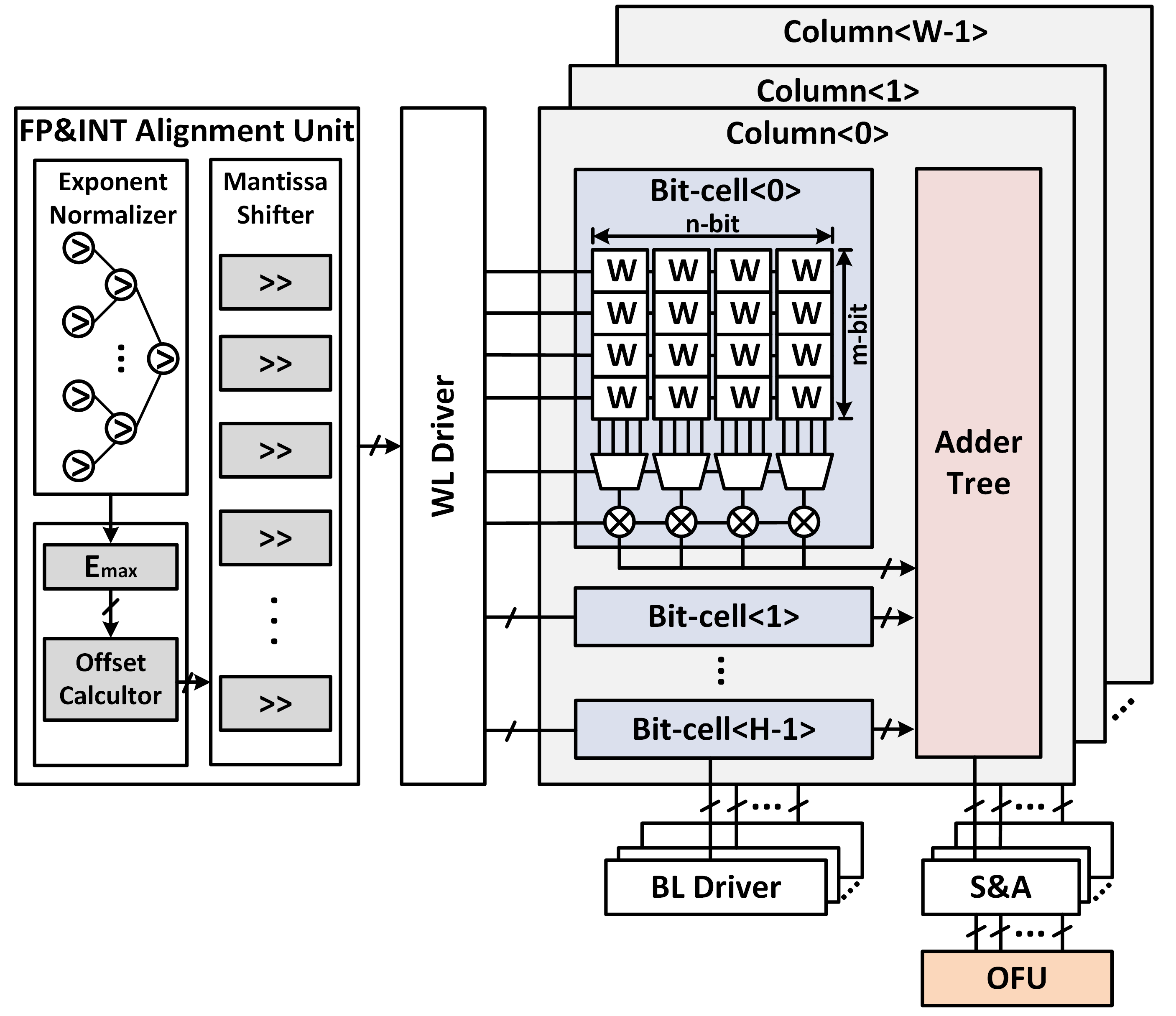}}
\caption{\small{Illustration of emerging DCIM architectures and key subcircuits.}}
\vspace{-3mm}
\label{fig_dcim_pre}
\end{figure}

DCIM has recently gained traction in AI applications, including edge computing, cloud computing, and on-chip training. A series of DCIM macro research from TSMC \cite{chih202116, fujiwara20225, mori20234nm, fujiwara202434} highlights the high efficiency and process scalability of DCIM. As shown in Figure \ref{fig_dcim_pre}, during the MAC operation, weights are stored in SRAM cells via bitline (BL) drivers. Input activations are fed into the array bit-serially through wordline (WL) drivers. Partial sums are accumulated using an adder tree and stored in a shift adder. To support multiple integer (INT) and floating-point (FP) precisions, the FP and INT alignment unit converts FP data into INT format through comparison and shifting, and then the output fusion unit fuses results across columns. Several DCIM macros have implemented memory-compute ratio (MCR)-aware designs to enhance on-macro memory density and facilitate efficient weight updates and MAC operations through specialized multiplier and multiplexer circuits.

\subsection{DCIM Subcircuits}

A typical DCIM macro consists of seven key subcircuits: FP\&INT Alignment Unit, WL/BL Driver, Memory Cell, Bitwise Multiplier and Multiplexer, Adder Tree, Shift and Adder (S\&A), and Output Fusion Unit (OFU). Each subcircuit can have various designs tailored to specific goals. Below is an analysis of the types and functionalities of these subcircuits.

\textbf{FP\&INT Alignment Unit}: This unit translates floating-point format data to integer format as required by the DCIM macro through a comparator tree and shifters\cite{tu2022redcim}. The complexity of this unit depends on the combination of required FP precisions. 

\textbf{WL/BL Driver}: The WL driver feeds input data and SRAM write/read signals into the DCIM array, while the BL driver writes weights into the SRAM array for updating. The power and size of the WL/BL driver depend on the array dimensions.

\textbf{Memory Cell}: Typically, a 6T SRAM cell is used for storage, with additional transistors for multiplication or read selection. Special designs may use an 8T D-latch cell for robust read and write \cite{mori20234nm}, or a 12T OAI gate-based cell for design feasibility \cite{lou2022all}. Novel hybrid designs \cite{ReSCIM, CREAM, TL-nvSRAM-CIM} integrate ReRAM and SRAM, leveraging ReRAM for storage while utilizing SRAM for ReRAM data reading and MAC computations.

\textbf{Multiplier and Multiplexer}: Various methods exist for multiplication and SRAM selection: (1) AutoDCIM uses a 1T passing gate as the multiplexer, which is area-efficient but has a voltage drop that affects power and latency. (2) \cite{mori20234nm} uses an OAI22 gate as a fused multiplier and multiplexer, saving wiring overhead but becoming less scalable when the memory-compute ratio (MCR) exceeds 2. (3) \cite{fujiwara20225} uses a 2T transmission gate for selection and a NOR gate for multiplication, which is a commonly adopted approach.
 
\textbf{Adder tree}: The adder tree accumulates partial sums across rows of a column and is a major power consumer. Typically composed of multi-stage signed ripple-carry adders (RCAs), it can be logically complex and reduce throughput. Some designs use bitwise compressor-based approximations for better efficiency and throughput, though this requires extra retraining. A novel design \cite{you20241} uses a 4-2 compressor as a 5-3 carry-save adder with an RCA for final stage accumulation, which is fast, power efficient and friendly for precision-configuration, while the 4-2 compressor is slow and there is lack of latency balance for signal paths.

\textbf{S\&A}: This unit accumulates the partial sums of bit-serial input. Its complexity is related to the input bit-width and the height of the DCIM macro.

\textbf{OFU}: For multi-precision-oriented reconfigurability, the OFU adds the outputs of the S\&As stage by stage, from lower bit-width to higher bit-width \cite{tu2022redcim}.

By analyzing and categorizing these subcircuits, DCIM design can be modularized into blocks and assembled through synthesis.

\section{SynDCIM: Performance-Aware DCIM Compiler}

\begin{figure}[tp]
\centerline{\includegraphics[width=\columnwidth]{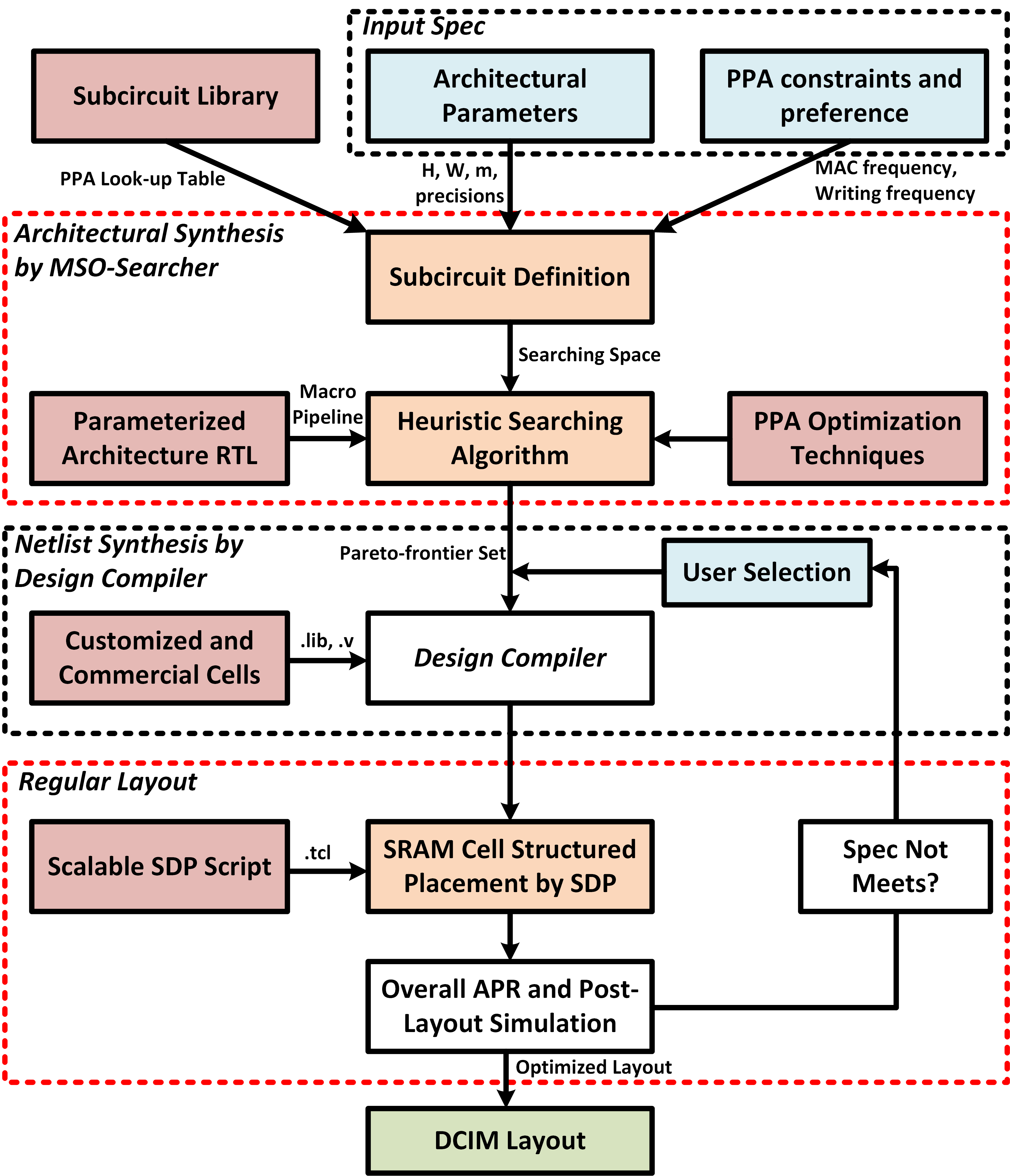}}
\caption{\small{Overall framework of the proposed performance-aware DCIM compiler, SynDCIM.}}
\vspace{-3mm}
\label{fig_syn_dcim_overall}
\end{figure}

\subsection{Overall Framework}

As depicted in Figure \ref{fig_syn_dcim_overall}, we propose SynDCIM, a performance-aware DCIM compiler with multi-spec-oriented subcircuit synthesis. SynDCIM is an end-to-end framework that generates a DCIM macro from architecture and performance specifications, optimizing for throughput, latency, power and area. Beyond the classic DCIM architecture, SynDCIM employs a bit-wise carry-save-adder (CSA) and configures the output fusion unit for multi-bit weight precisions. The SynDCIM compiler consists of three main components: (1) a subcircuit library, (2) a multi-spec-oriented (MSO) searcher, and (3) synthesis and SDP-based automatic place and routing (APR).

SynDCIM takes architectural parameters such as dimensions, FP\&INT precisions, MCR, and performance constraints including MAC frequency, weight updating frequency, and power-performance-area (PPA) preferences as input specifications.

To expand the design space and generate a DCIM macro at the Pareto frontier, we build a subcircuit library that integrates a PPA lookup table (LUT) of optimized subcircuits under different circuit topologies, dimensions, and timing constraints.

Once the input specifications are determined, the searcher defines the configurations and constraints of each subcircuit according to the architecture parameters, forming a search space of selected subcircuits. Based on this search space, the searcher executes the architectural synthesis for the DCIM macro and evaluates and optimizes the PPA using a heuristic search algorithm, ensuring that the macro meets performance constraints. A series of DCIM designs at Pareto frontiers are generated for subsequent synthesis and APR.

Among these optimal designs, one is finally selected by the user for implementation. The architecture RTL, subcircuit RTL and netlist are synthesized by design compiler and generated into a netlist for APR. At the APR stage, we designed a scalable SDP script for regular SRAM place and uniform routing, and the adders are placed next to the regular SRAM column by the EDA tool. Final PPA data is evaluated through post-simulation after design rule check (DRC) and layout versus schematic (LVS) verification.

\begin{figure}[tp]
\centerline{\includegraphics[width=\columnwidth]{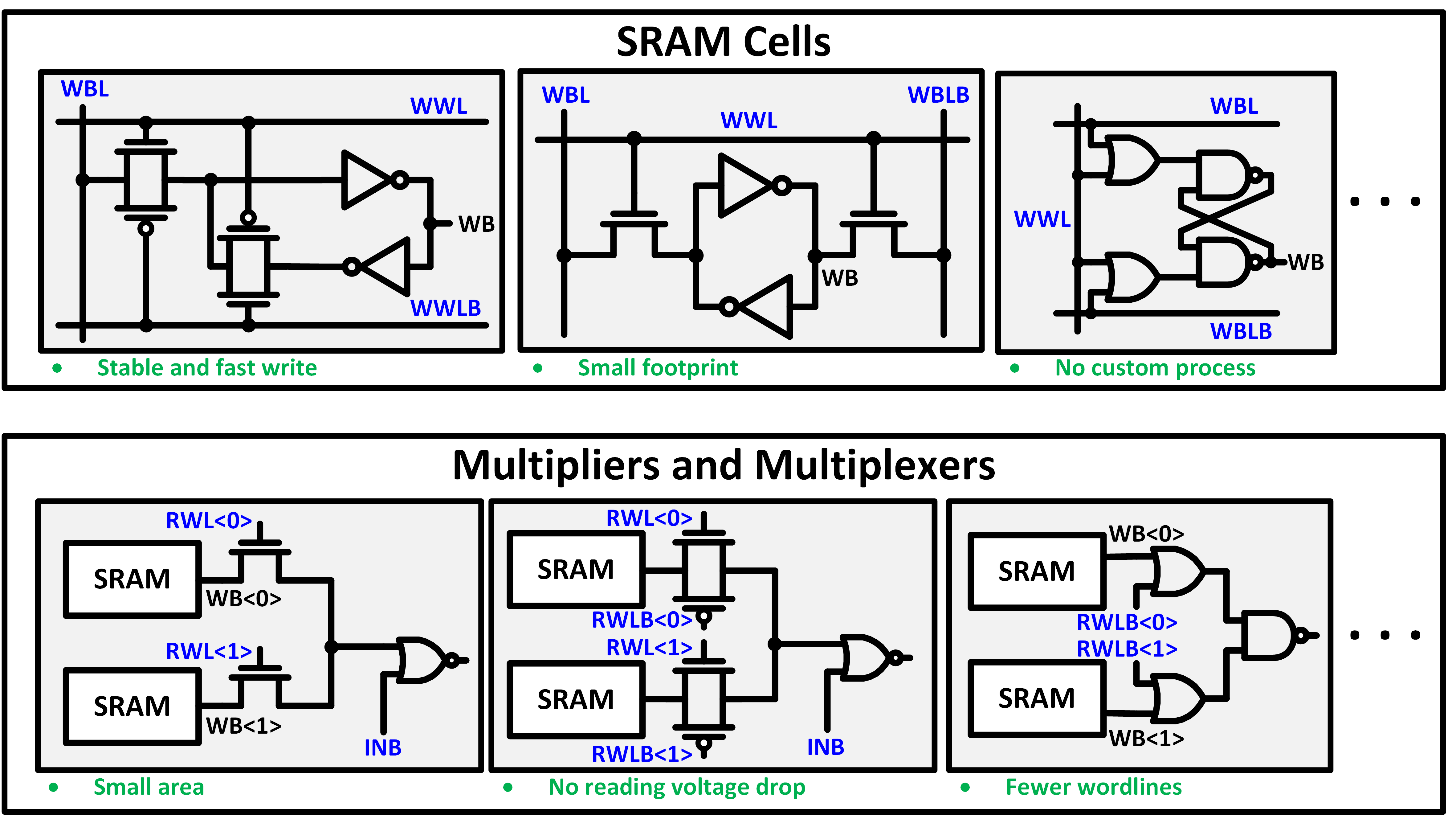}}
\caption{\small{Illustration of the proposed DCIM subcircuit library for synthesis.}}
\vspace{-3mm}
\label{fig_syn_dcim_subcircuit}
\end{figure}

\begin{figure}[tp]
\centerline{\includegraphics[width=\columnwidth]{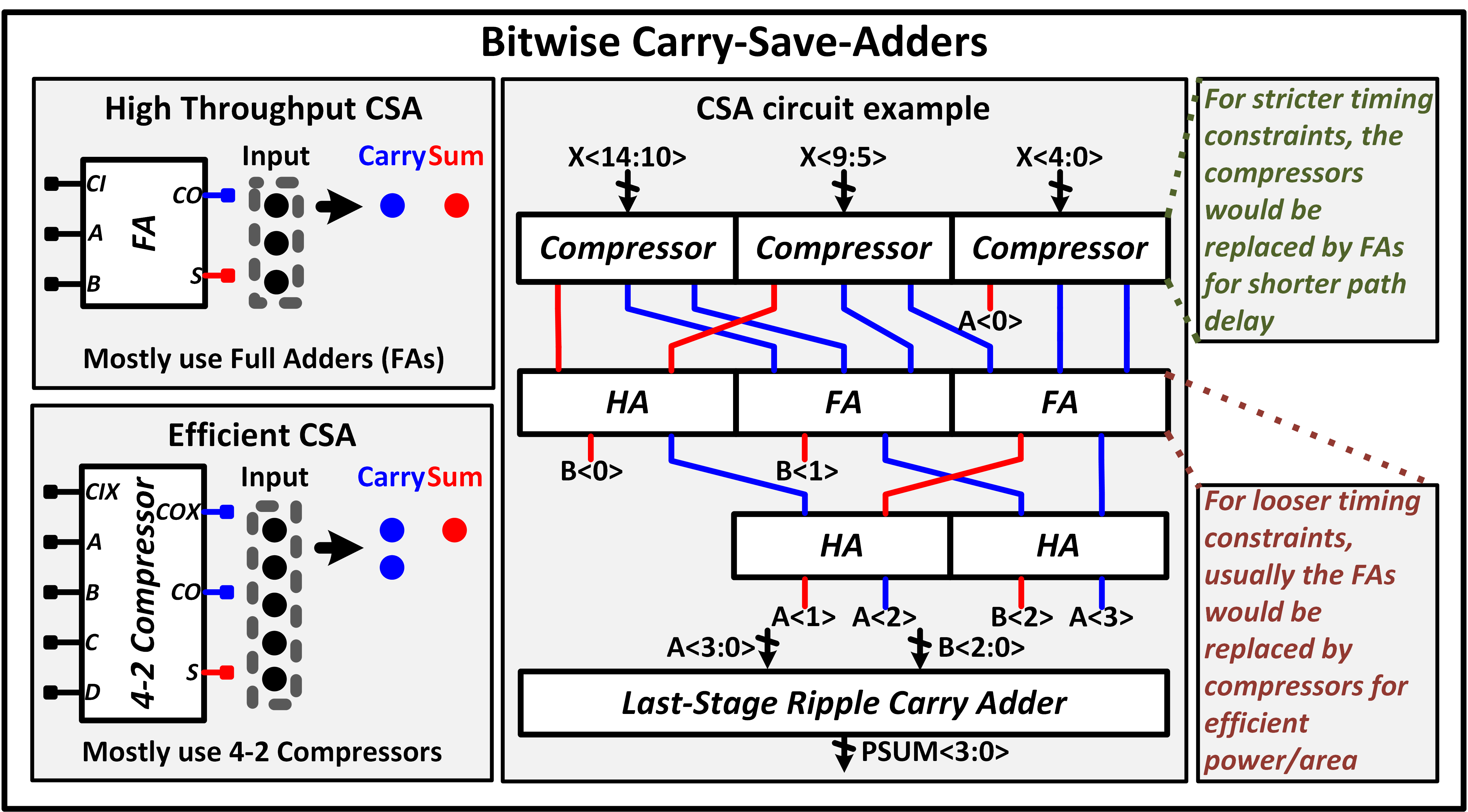}}
\caption{\small{Proposed bit-wise carry-save-adders: principle and circuit details.}}
\vspace{-3mm}
\label{fig_syn_dcim_csa_detail}
\end{figure}

\subsection{Subcircuit Library for Synthesis}

For the seven types of subcircuits, we build a Subcircuit Library (SCL) that includes PPA lookup tables (LUTs) for subcircuits of various topologies, dimensions, and timing constraints.

As shown in Figure \ref{fig_syn_dcim_subcircuit}, for customized circuits like SRAM cells, multipliers, and multiplexers, we design the layout and obtain PPA data through custom cell characterization flow, making them standard cells for integration into the digital flow. We also design different topologies and combinations for these components to meet specific requirements.

We design a series of bit-wise carry-save adders (CSAs) tailored for different PPA preferences, using a mix of 4-2 compressors, full adders, and half adders. Previous work has shown that bit-wise 4-2 compressor-based CSAs are faster, smaller, and more energy-efficient than conventional signed RCA-based adder trees. However, they lack optimization due to unbalanced paths, and while 4-2 compressors are power- and area-efficient, they are relatively slower than full adders. To address this, we propose a mixed compressor- and full adder-based CSA, as shown in Figure \ref{fig_syn_dcim_csa_detail}. For loose timing constraints, we use more 4-2 compressors to minimize power and area consumption. For strict timing constraints, we replace 4-2 compressors with full adders to shorten the critical path, sacrificing power and area. Additionally, due to the different path delays of each cell port, the carry bit is faster than sum bits, presenting a timing optimization opportunity by reordering the connections between cells.

We build parameterized RTL templates for other purely digital subcircuits, including the FP/INT alignment unit, WL/BL buffer, S\&A, and OFU. Typical configurations are implemented into layouts and simulated for PPA data. The PPA data for other configurations can be estimated and scaled from synthesis data.

By integrating these subcircuits into the SCL, we enhance the flexibility and efficiency of the SynDCIM compiler, ensuring it can generate optimized DCIM macros tailored to various specifications and performance requirements.

\begin{figure}[tp]
\centerline{\includegraphics[width=\columnwidth]{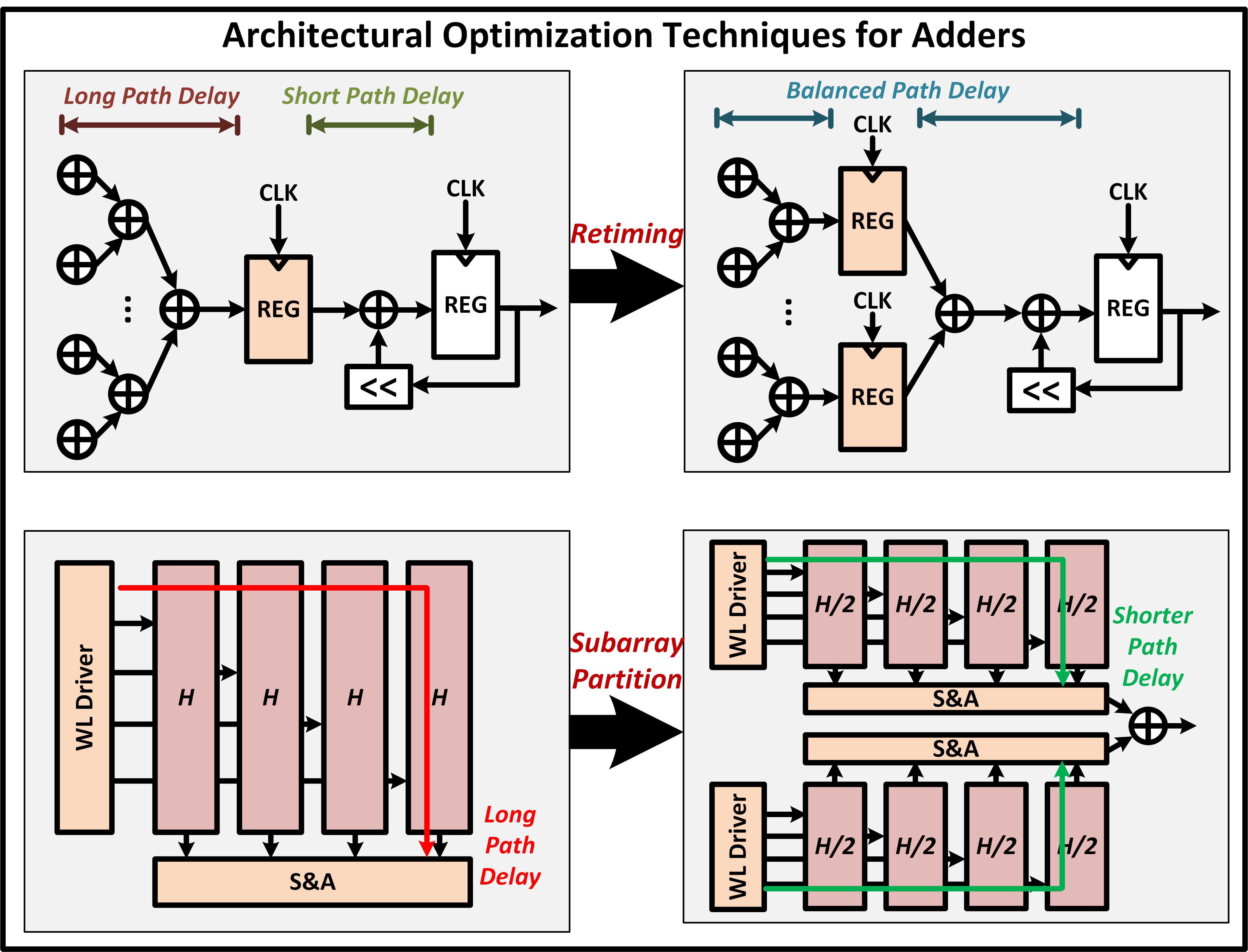}}
\caption{\small{Illustration of the proposed optimization techniques for adders.}}
\vspace{-3mm}
\label{fig_mso_adder}
\end{figure}

\subsection{Multi-Spec-Oriented Searching Algorithm}

As shown in Figure \ref{fig_mso_adder} and Algorithm \ref{alg:msc_search}, we propose a multi-spec-oriented searcher based on a heuristic search algorithm to optimize the DCIM macro at the architectural level. The searcher takes the input specifications and subcircuit library as inputs and synthesizes a series of optimized RTL/netlists of DCIM macros at the Pareto frontier, chosen based on PPA preferences.

When the input specifications are determined, we first define the configurations of each subcircuit based on these specifications, forming a search space. This search space contains selectable subcircuits with architecture parameters defined by the specifications, which can be directly synthesized by the searcher.

\begin{algorithm}
\caption{Heuristic Hierarchical Search Algorithm}
\centering
\includegraphics[scale=0.7]{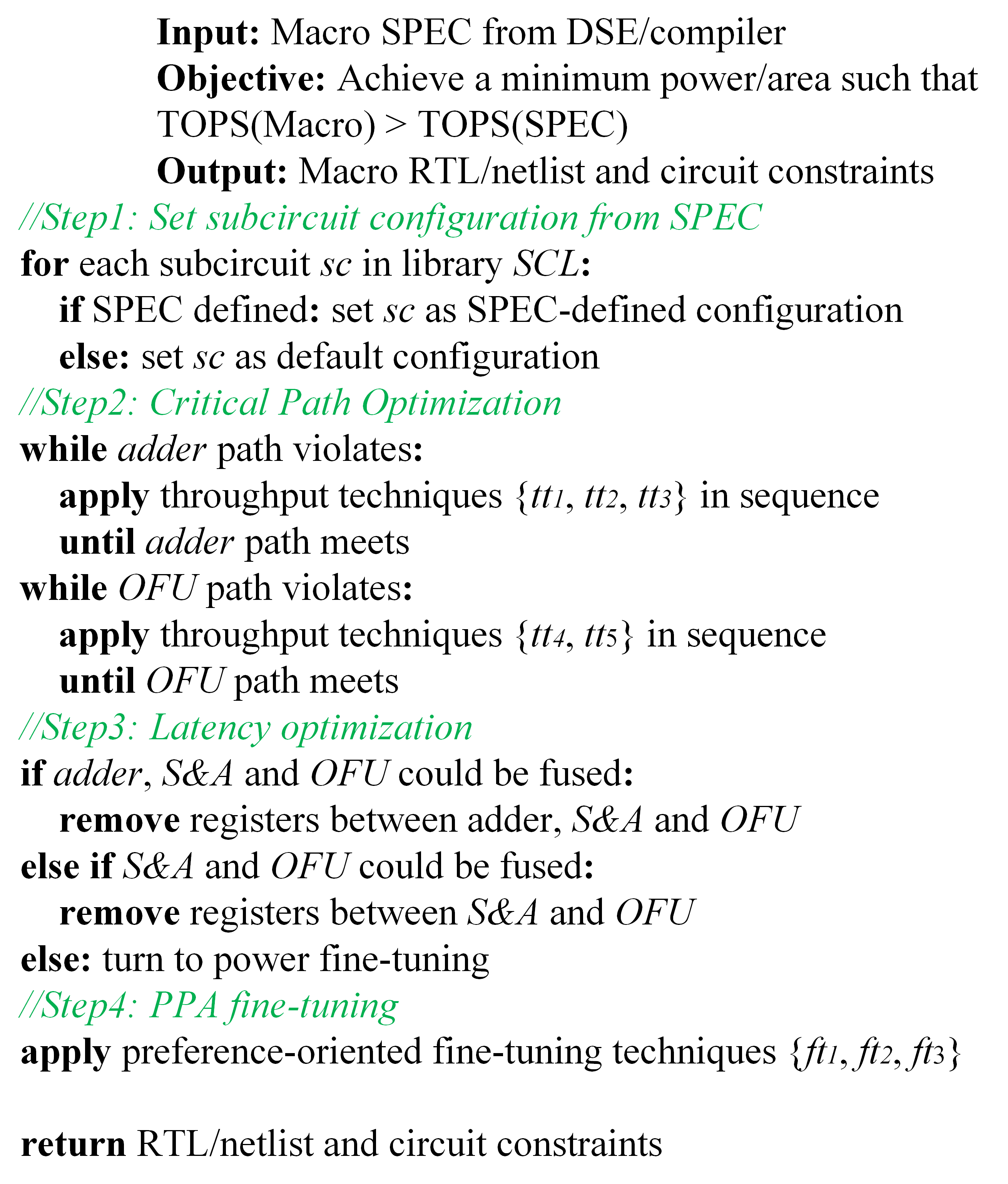}
\label{alg:msc_search} 
\end{algorithm}

Once the search space is ready, the searcher evaluates whether the critical paths of the MAC, including the WL driver, multiplier, adder tree, and OFU (which contains stages of additions), meet the timing constraints. For the MAC path, the searcher checks if faster adders are available in the SCL or performs retiming by moving the registers at the output of the adder to the front of the last RCA stage. If these fine-tuning techniques do not work, the searcher divides the column with height H into two columns with height H/2. Similarly, if the OFU does not meet the timing constraints, the searcher performs retiming by moving some combinational circuits to the S\&A. If retiming is insufficient, the searcher adds an extra pipeline stage to the OFU. 

After satisfying the basic timing requirements, the searcher optimizes the pipeline registers between the adder tree and S\&A, and between S\&A and OFU. If the combined path delay of neighboring combinational circuits still meets the timing constraints, the searcher removes the registers between them. Finally, fine-tuning optimization techniques for power or area are applied by substituting power/area-efficient subcircuits.

The synthesized and optimized design points are output as RTL and netlists for further selection, ensuring that the resulting DCIM macros are optimized for the specified performance constraints and PPA preferences.

\begin{figure}[tp]
\centerline{\includegraphics[width=\columnwidth]{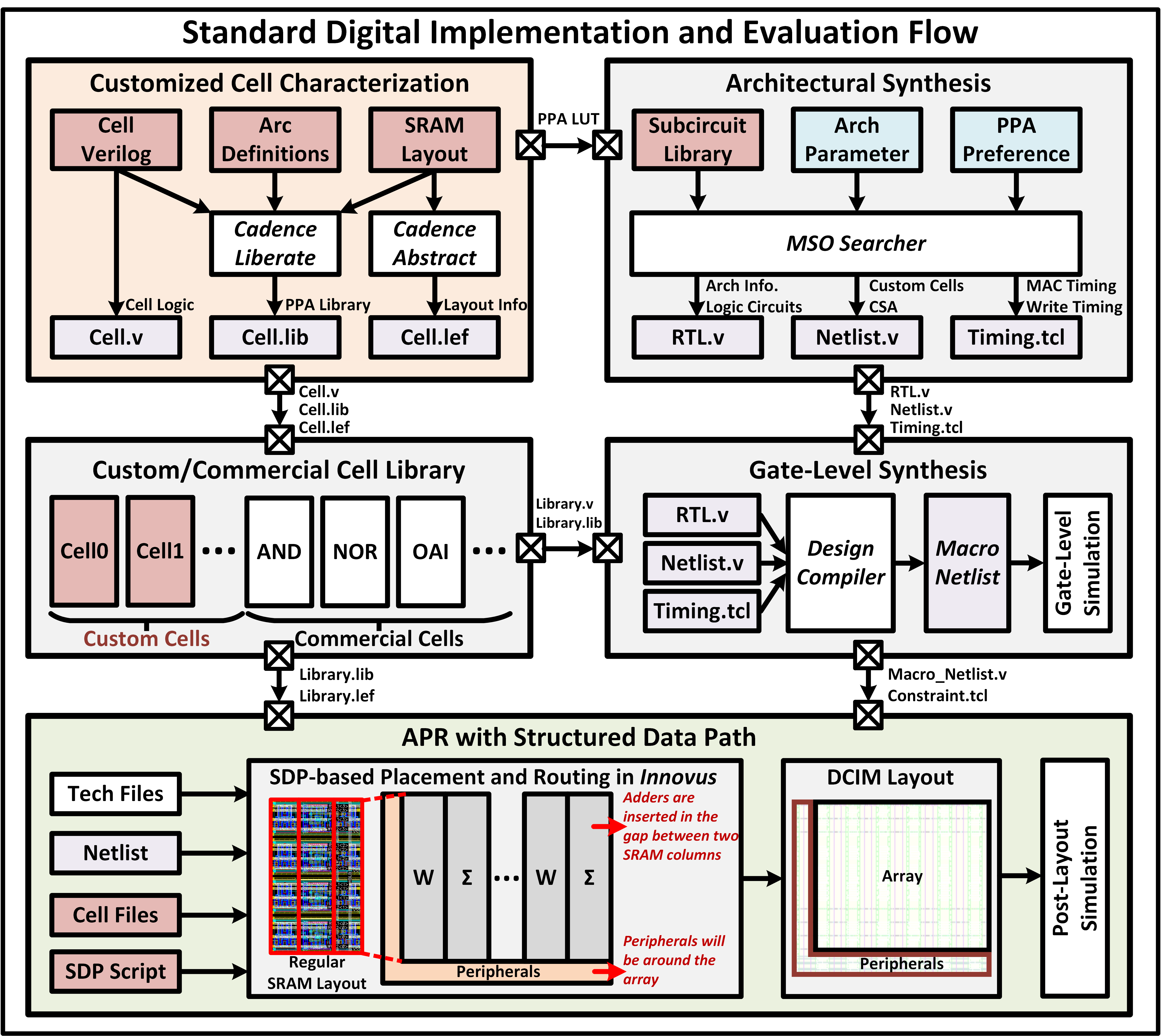}}
\caption{\small{The proposed implementation and evaluation flow of SynDCIM.}}
\vspace{-3mm}
\label{fig_impl_eval_flow}
\end{figure}

\subsection{Implementation and Evaluation Flow}

SynDCIM executes the layout implementation through a standard digital flow involving synthesis and APR, as shown in Figure \ref{fig_impl_eval_flow}. 
Due to the characterization of the customized cells, the LEF file (describing the GDS information) and the LIB file (providing timing, power, and area information) are compatible with standard cells, allowing integration into the standard digital flow. During placement by APR tools, cells may be scattered, affecting macro performance. To generate a regular layout, we adopt the structured data path (SDP) capability in Cadence Innovus with a scalable script. This enables structured cell placement and uniform routing by sourcing an SDP TCL script. After placing the SRAM cells using SDP, we fill the gaps between SRAM columns with adder cells and place the peripheral logic around the array using APR tools.
Following synthesis by EDA tools, we evaluate the PPA of the netlist through gate-level simulation to ensure it meets frontend requirements. After design rule check (DRC) and layout versus schematic (LVS) verification, we perform post-layout simulation to confirm that the DCIM macro meets the specified functionality and performance criteria.

\section{Experiments and Validation}

In this section, we conduct several detailed studies for SynDCIM and demonstrate its effectiveness with a fabricated test DCIM chip. For post-layout simulation experiments, the gate-level netlist is synthesized using Synopsys Design Compiler, APR is executed in Cadence Innovus, and both gate-level and post-layout simulations are performed with Synopsys PrimeTime. For silicon validation, we fabricate the SynDCIM-generated macro with a 40nm CMOS technology for measurement and test.

\subsection{Post-layout Evaluations}

\textbf{Energy Efficiency with Different Precisions} In this experiment, we generate four DCIM macros with dimensions ranging from 32x32 to 256x256 and evaluate their power consumption for executing MAC operations in INT4/8, FP8, and BF16 formats. We explore the scaling trend and the overhead of the configurable FP/INT Alignment Unit and OFU.

\begin{figure}[tp]
\centerline{\includegraphics[width=1.1\columnwidth]{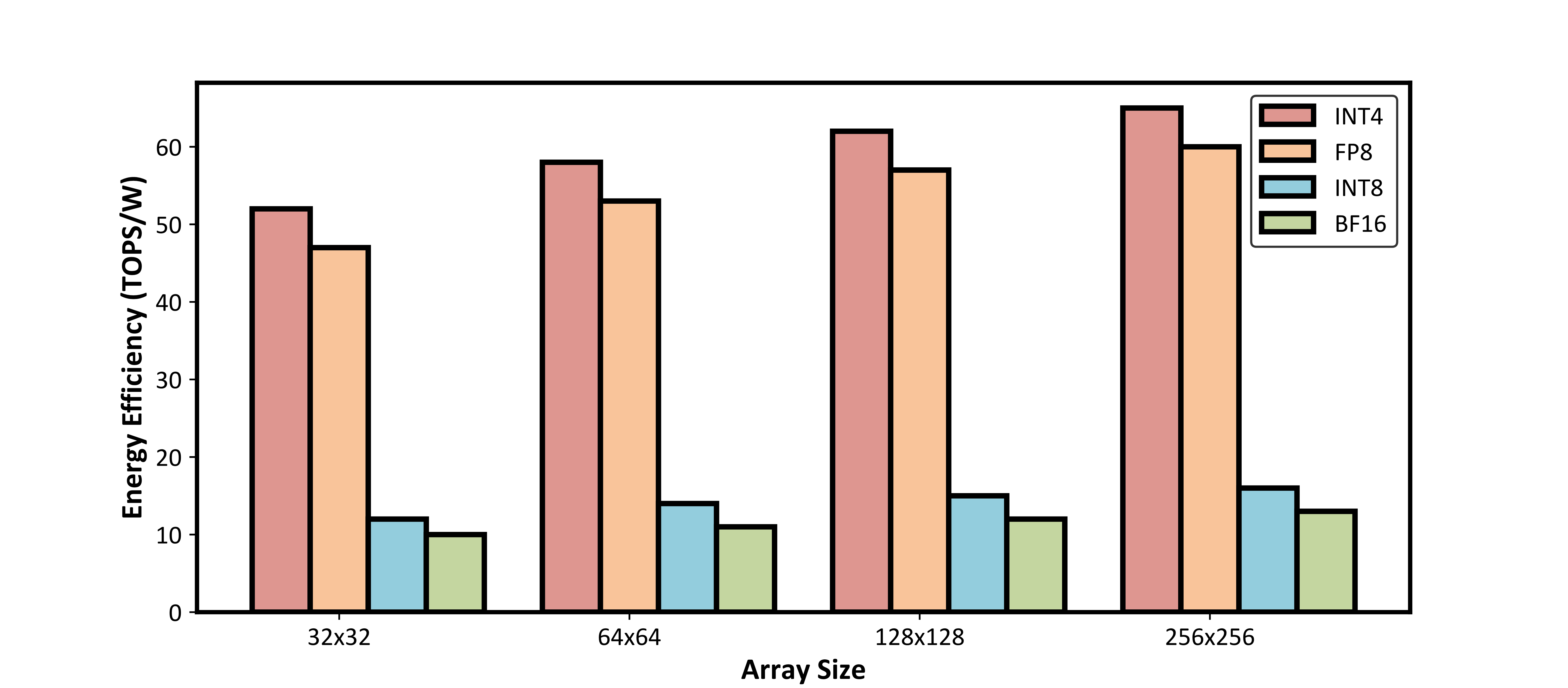}}
\caption{\small{Post-layout energy efficiency of SynDCIM-generated DCIM macros across different precisions and dimensions.}}
\vspace{-3mm}
\label{fig_result_ef_prec}
\end{figure}

\begin{figure}[tp]
\centerline{\includegraphics[width=\columnwidth]{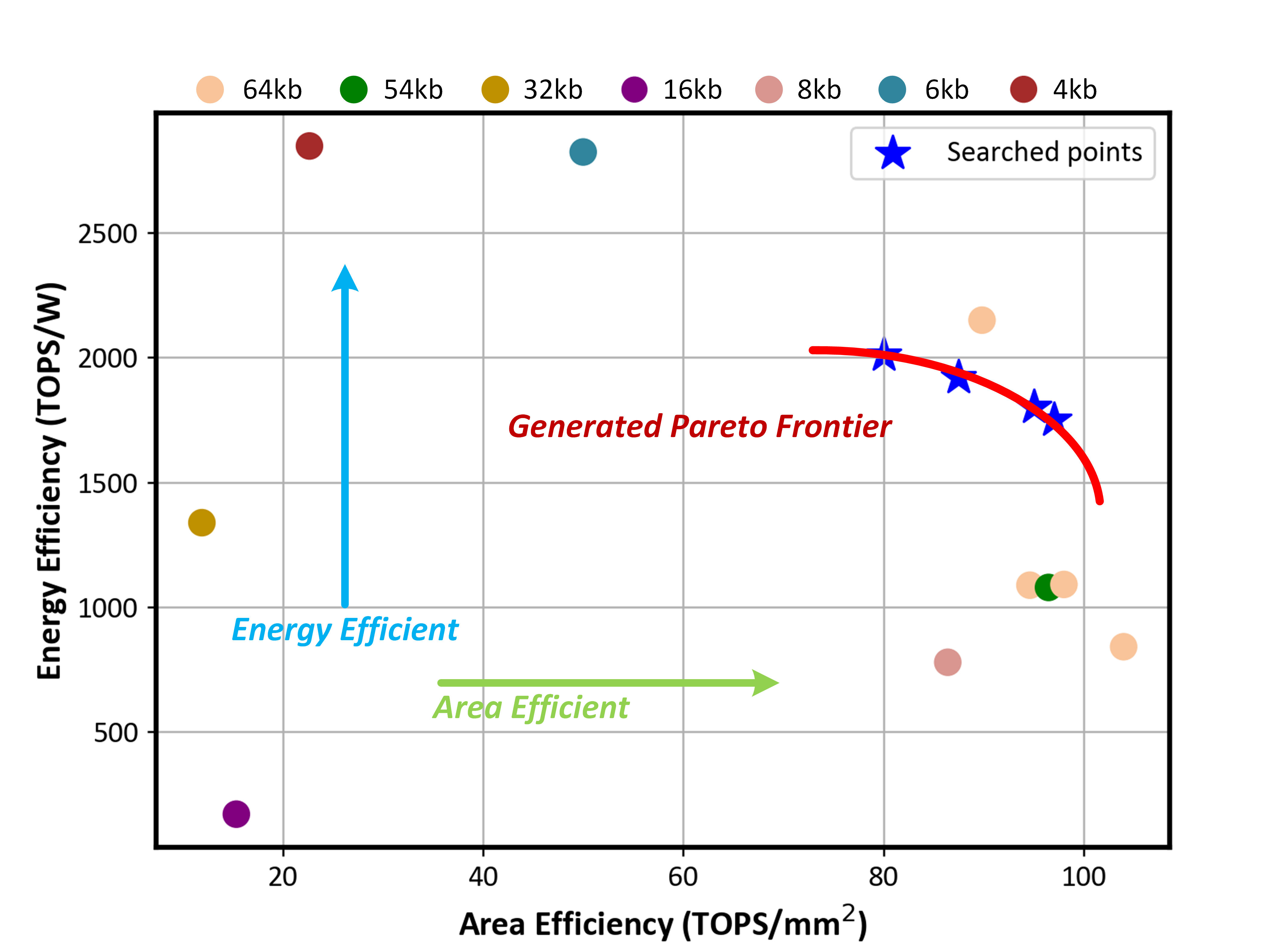}}
\caption{\small{Pareto-Frontier of SynDCIM generated DCIM macros and comparisons.}}
\vspace{-3mm}
\label{fig_result_pf}
\end{figure}

As shown in Figure \ref{fig_result_ef_prec}, as the dimensions of the DCIM array increase, energy efficiency also improves due to lower amortized peripheral overhead per bit and higher energy efficiency of the CSA. However, due to the FP/INT alignment overhead, FP8 and BF16 consume around 10\% and 20\% more power than INT4 and INT8, respectively.

\textbf{Searched and Generated Pareto-Frontier} For a given specification, the MSO Searcher generates a series of design points at the Pareto frontier. We define the specifications as follows: H=W=64, MCR=2, precisions = INT4/8, FP4/8, MAC and weight update frequency @0.9V = 800MHz. The generated and implemented designs are shown and labeled in Figure \ref{fig_result_pf}.
The top designs are energy-efficient with low power, and the right designs are area-efficient with small area or high throughput. The generated designs are partly biased towards energy efficiency and partly towards area efficiency, to be finally chosen based on defined PPA preferences or user selection. Four typical designs are selected and implemented into layouts, forming a Pareto frontier. Our generated design points balance power efficiency and area efficiency.

\subsection{Silicon Validation}

\begin{figure}[tp]
\centerline{\includegraphics[width=\columnwidth]{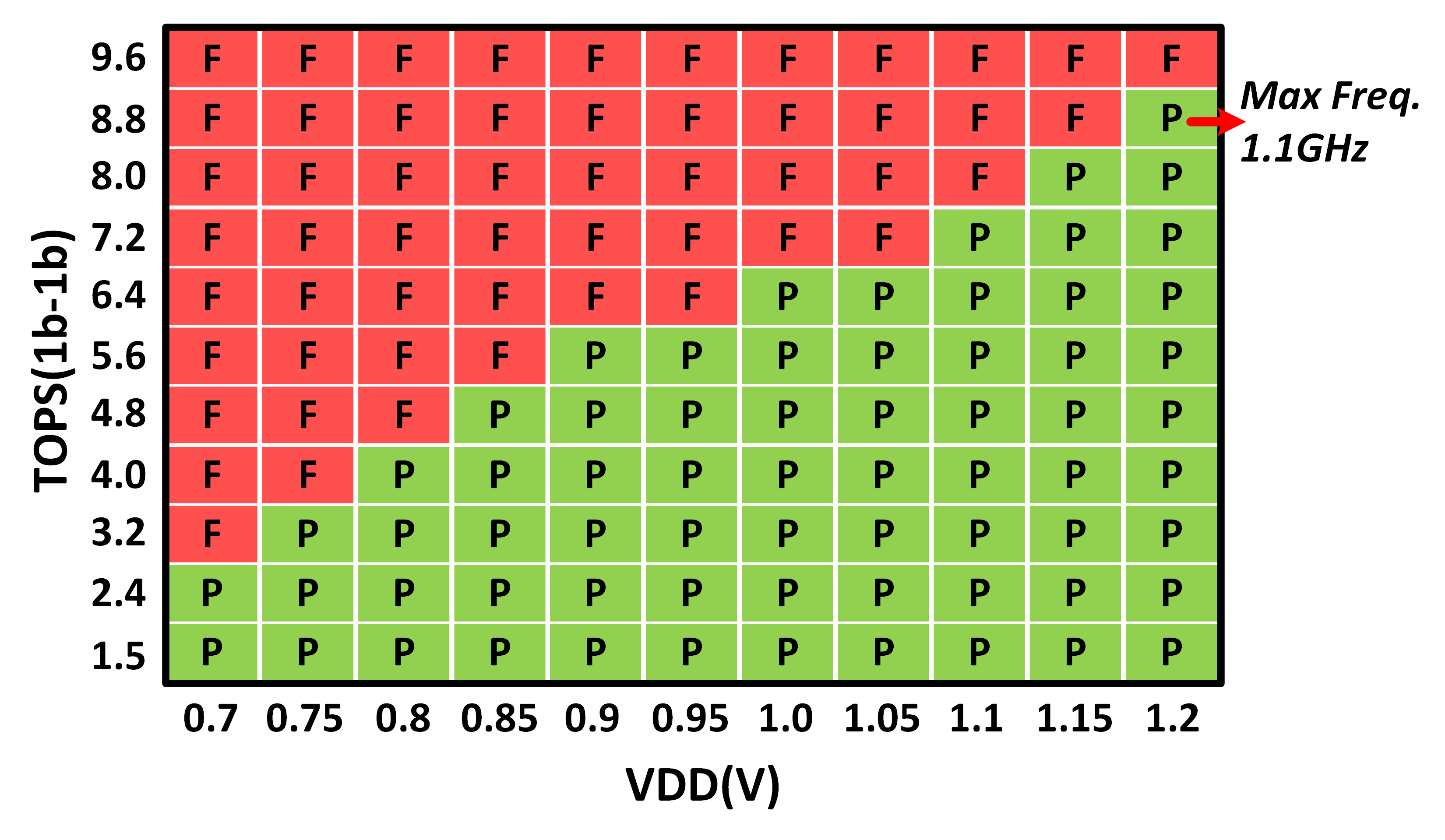}}
\vspace{-3mm}
\caption{\small{Shmoo plot of the test DCIM macro generated by SynDCIM.}}
\vspace{-3mm}
\label{fig_result_shmoo}
\end{figure}

A test chip for a 64×64, MCR=2 DCIM macro with INT1/2/4/8 and FP4/8, generated by the proposed SynDCIM compiler, has been fabricated using 40nm CMOS technology.

Figure \ref{fig_result_shmoo} shows the shmoo plot of the test chip, indicating an ultra-high frequency of 1.1GHz and throughput of 9 TOPS at a supply voltage of 1.2V. When operating at 0.7V for higher efficiency, the maximum frequency achieves 300MHz, enabling efficient processing with high throughput.

\begin{figure}[tp]
\centerline{\includegraphics[width=0.5\columnwidth]{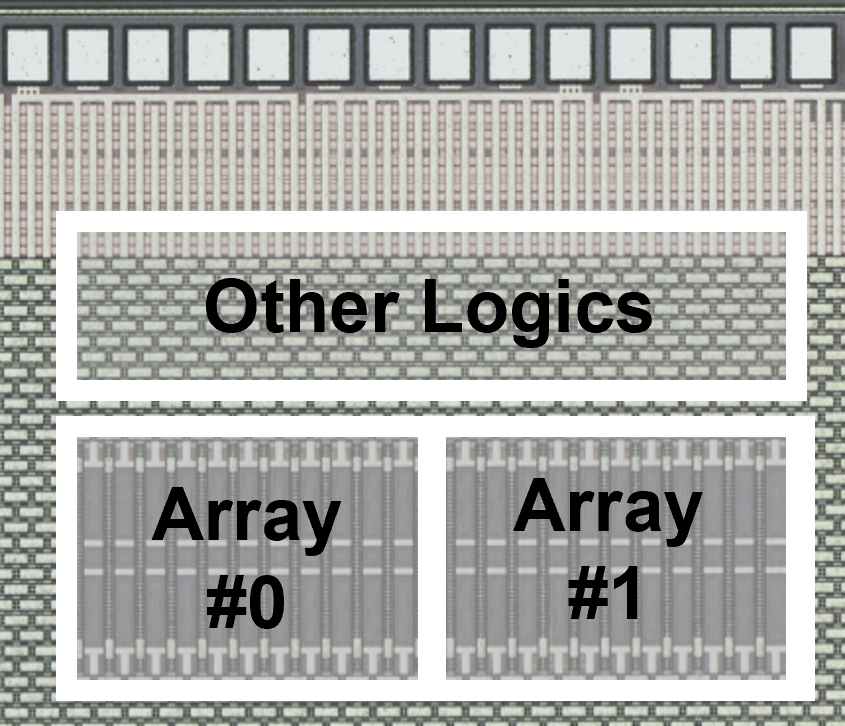}}
\caption{\small{Die photo of the fabricated 40nm test DCIM macro chip generated by SynDCIM.}}
\vspace{-3mm}
\label{fig_result_chip}
\end{figure}

Figure \ref{fig_result_chip} shows a micrograph of the test chip fabricated in a 40nm CMOS technology node, with eight DCIM macros implemented on a chip. The area of one DCIM macro is 0.112mm$^{2}$ (455×246um$^{2}$).

\begin{table}[t]
\caption{\textcolor{black}{Comparison of the Test DCIM Macro Chip Generated by SynDCIM with State-of-the-Art DCIM Designs.}}
\centering
\includegraphics[scale=0.6]{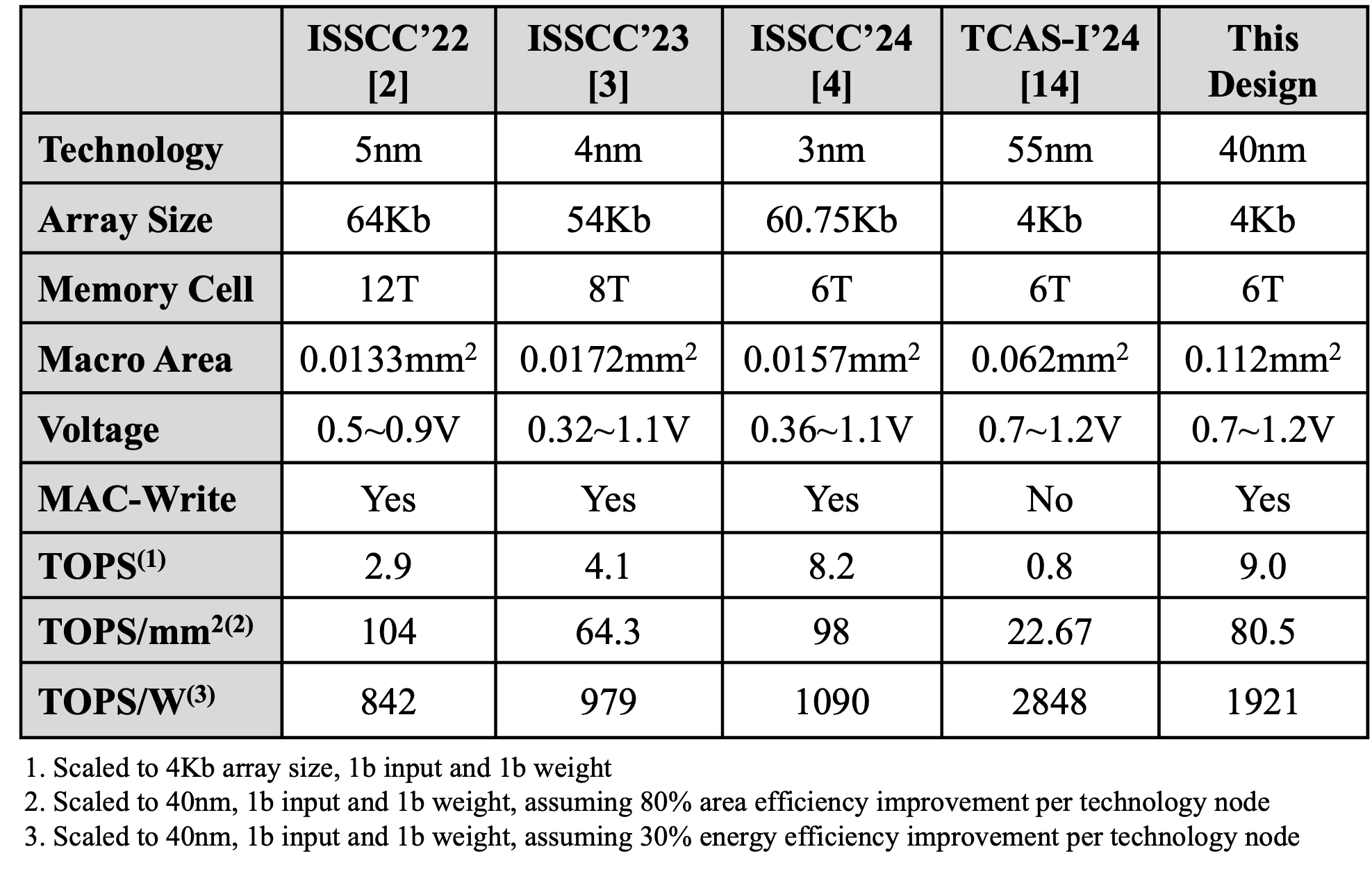}
\vspace{-3mm}
\label{tab:comparison_dcim_macro} 
\end{table}

Table \ref{tab:comparison_dcim_macro} shows the measured performance of our test chip with input sparsity of 12.5\% and weight sparsity of 50\% in INT4 at 25°C. Measurements show an ultra-high energy efficiency of 1921 TOPS/W and a comparable area efficiency of 80.5 TOPS/mm$^{2}$ (scaling to 1b-1b) compared with state-of-the-art designs.

\section{Conclusion}

This paper introduces SynDCIM, a pioneering performance-aware DCIM compiler designed to meet the diverse execution requirements of AI applications. SynDCIM offers an agile solution that integrates user-defined performance metrics into the design process with a comprehensive subcircuit library and multi-spec-oriented synthesis. Our validation of SynDCIM using a silicon-verified DCIM macro demonstrates its effectiveness and highlights its superiority over existing solutions in the field. This work not only bridges the gap between user expectations and DCIM macro generation processes but also paves the way for greater flexibility and efficiency in deploying DCIM technologies across various AI applications and acceleration scenarios.

\clearpage

\bibliographystyle{unsrt}
\bibliography{reference}

\end{document}